\documentclass[aps,prb,longbibliography,showpacs,floatfix,superscriptaddress,twocolumn]{revtex4-1}

\usepackage{graphicx,color}
\usepackage[justification=RaggedRight]{caption}
\usepackage{subcaption}
\usepackage{amsfonts}
\usepackage[figuresright]{rotating}  
\usepackage{amssymb}
\usepackage{amsmath}
\usepackage{mathtools}
\usepackage{psfrag}
\usepackage{multirow}
\usepackage{tabularx}
\usepackage{textcomp}
\usepackage{units}
\usepackage{lipsum}
\usepackage{soul}
\usepackage{titlesec}
\usepackage{times}
\usepackage{hyperref}
\usepackage{enumitem}
\usepackage{ulem}
\usepackage{braket}

\newcommand{\bk}{\boldsymbol{k}} % bold k
\newcommand{\mc}{\mathcal}
\definecolor{green}{rgb}{0, 1, 0}
\newcommand\identity{1\kern-0.25em\text{l}}
\graphicspath{{figures/}}
\begin{document}

\title{Topological skyrmion semimetals}

\author{Shu-Wei Liu}
\affiliation{Max Planck Institute for the Physics of Complex Systems, N{\"o}thnitzer Stra{\ss}e 38, 01187 Dresden, Germany}
\affiliation{Max Planck Institute for Chemical Physics of Solids, N{\"o}thnitzer Stra{\ss}e 40, 01187 Dresden, Germany}
\author{Joe H. Winter}
\affiliation{Max Planck Institute for the Physics of Complex Systems, N{\"o}thnitzer Stra{\ss}e 38, 01187 Dresden, Germany}
\affiliation{Max Planck Institute for Chemical Physics of Solids, N{\"o}thnitzer Stra{\ss}e 40, 01187 Dresden, Germany}
\affiliation{SUPA, School of Physics and Astronomy, University of St.\ Andrews, North Haugh, St.\ Andrews KY16 9SS, UK}
\author{Ashley M. Cook}
\affiliation{Max Planck Institute for the Physics of Complex Systems, N{\"o}thnitzer Stra{\ss}e 38, 01187 Dresden, Germany}
\affiliation{Max Planck Institute for Chemical Physics of Solids, N{\"o}thnitzer Stra{\ss}e 40, 01187 Dresden, Germany}

\begin{abstract}
We introduce topological skyrmion semimetal phases of matter, characterized by bulk electronic structures with topological defects in ground state observable textures over the Brillouin zone (BZ), rather than topological degeneracies in band structures. We present and characterize toy models for these novel topological phases, focusing on realizing such topological defects in the ground state spin expectation value texture over the BZ. We find generalized Fermi arc bulk-boundary correspondences and chiral anomaly response signatures, including Fermi arc-like states which \textit{do not} terminate with topological band structure degeneracies in the bulk, but rather with topological defects in the spin texture of bulk \textit{insulators}. We also consider novel boundary conditions for topological semimetals, in which the 3D bulk is mapped to a 2D bulk plus 0D defect. Given the experimental significance of topological semimetals, our work paves the way to broad experimental study of topological skyrmion phases and the quantum skyrmion Hall effect.
\end{abstract}

\maketitle

Topological semimetals are essential to experimental study of topological condensed matter physics, given that some are realized through breaking of symmetries---rather than symmetry-protection---as in the case of the Weyl semimetal (WSM)\cite{Armitage2018, sun2015}.  
These three-dimensional (3D) phases of matter are realized by breaking time-reversal or spatial inversion symmetry, exhibiting topologically-robust two-fold band structure degeneracies\cite{soluyanov2015, xu2015'', Teo2008, yang2014, young2012} with distinctive consequences such as Fermi arc surface states\cite{wan2011, balents2011, vishwanath2015, Hasan2017, huang2015, Lv2015, xu2015, chan2016} and the chiral anomaly\cite{nielsen1983, song2013, parameswaran2014, huang2015', liang2016}. 

These topological phases are associated with mappings to the space of projectors onto occupied states, as are all other previously-known topological phases descending from the ten-fold way classification scheme~\cite{ryu2010, schnyder2008}. Recently-introduced~\cite{cook2023} topological skyrmion phases (TSPs) of matter, however, broadly generalize these concepts by considering mappings to the space of observable expectation values, $ \langle \mathcal{O} \rangle$. While some TSPs have already been introduced~\cite{cook2023,cook2023QSkHE, liu2023, florescalderon2023}, the full set of these phases of matter is currently unknown and requires generalization of the four-decade-old framework~\cite{laughlin1983} of the quantum Hall effect to that of the quantum skyrmion Hall effect (QSkHE)~\cite{cook2023QSkHE}.
 
We introduce the topological skyrmion semimetals (TSSs) in this work, both to broadly generalize known topological semimetals and to facilitate the search for TSPs and the QSkHE in experiments. We first present recipes for constructing toy models inspired by Weyl semimetals, and then characterize bulk electronic structure, finding a bulk-boundary correspondence yielding generalizations of Fermi arc surface states, as well as a generalization of the chiral anomaly.  Notably, we construct a three-band Bloch Hamiltonian toy model for a TSS, which exhibits generalized Fermi arc states for a bulk \textit{insulator}, due to $\mc{Q}$ changing by a type-II topological phase transition~\cite{cook2023}, which occurs without the closing of the minimum direct bulk energy gap and while respecting the symmetries protecting the topological phase and maintaining fixed occupancy of bands, in effectively non-interacting systems. Our work is therefore a foundation for broad generalization of concepts of topological semimetals \textit{and insulators}.

%We introduce topological skyrmion semimetals first by constructing minimal four-band Bloch Hamiltonians realizing such topology, and later generalize the construction by considering richer three-band models with lower symmetry. Intuition and motivation for our approach derives from construction of Weyl semimetal Hamiltonians as a stack of Chern insulators, each defined over a 2D BZ in terms of momenta $k_x$ and $k_y$, for instance, along a third momentum axis $k_z$. 

\textit{Minimal model \textemdash}  We first consider a minimal two-band Bloch Hamiltonian for a Weyl semimetal, constructed from the Qi-Wu-Zhang (QWZ) model for a Chern insulator defined on a square lattice~\cite{qi2006} with additional dependence on a third momentum component, $k_z$, as $h(\bk) =  \sin{k_x} \sigma_x + \sin{k_y} \sigma_y + \left( 2-\cos{k_x}-\cos{k_y}+\gamma -\cos{k_z} \right) \sigma_z$, where $\sigma_i$ are the Pauli matrices, $\bk = \left( k_x,k_y,k_z\right)$ is momentum, and $\gamma$ is a constant. $h(\bk)$ realizes a Weyl semimetal phase for values of $\gamma$ such that the Chern number for the lower band of the model at fixed $k_z$ changes from one integer value to another across at least two values of $k_z$, with Weyl nodes realized as topologically-protected band-touching points at these values of $k_z$ required by the change in Chern number. We then construct the four-band model for a topological skyrmion phase relevant to centrosymmetric superconductors similarly to past work~\cite{liu2023}, in terms of the two-band WSM Hamiltonian $h(\bk)$ and its generalized particle-hole conjugate~\cite{LiuHopf2017, cook2023, liu2023} as
\begin{equation}
H(\bk) = \begin{pmatrix}
h(\bk) & \Delta_t(\bk)\\
\Delta^{\dagger}_t(\bk) & -h^{*}(\bk)
\end{pmatrix},
\label{eqn: minimal 4-band Weyl}
\end{equation}
where $\Delta_t(\textbf{k})$ is an additional spin triplet pairing term considered in previous work~\cite{cook2023, liu2023}, which takes the form
\begin{equation}
    \Delta_t(\bk) = i\Delta_0 \left(\boldsymbol{d}(\textbf{k}) \cdot \boldsymbol{\sigma}\right)\sigma_y.
\end{equation}
Here, $\Delta_0$ is the pairing strength and the $\boldsymbol{d}$-vector of the spin-triplet pairing term is taken to be $\boldsymbol{d}(\bk) =\sin(k_y) \hat{x} - \sin(k_x)\hat{y}$ for this example. This choice of  $\boldsymbol{d}$-vector has previously been proposed as characterizing Sr$_2$RuO$_4$ in the high-field phase~\cite{ueno2013}. 

For each value of $k_z$, we characterize topology of the corresponding $2$D submanifold of the BZ with two topological invariants, the total Chern number of occupied bands, $\mc{C}$, and the topological charge of the ground state spin expectation value texture over the BZ, or skyrmion invariant $\mc{Q}$, expressed in terms of the normalized ground-state spin expectation value $\langle \boldsymbol{\hat{S}}(\boldsymbol{k}) \rangle$ as in past work~\cite{cook2023, cook2023QSkHE, liu2023, florescalderon2023} as
\begin{equation}
    \mc{Q} = \frac{1}{4\pi} \int d\boldsymbol{k} \left[\boldsymbol{\hat{S}}(\boldsymbol{k}) \cdot \left(\partial_{k_x} \boldsymbol{\hat{S}}(\boldsymbol{k}) \times \partial_{k_y} \boldsymbol{\hat{S}}(\boldsymbol{k}) \right) \right].
\end{equation}
We may therefore also interpret Eq.~\ref{eqn: minimal 4-band Weyl} as a stack of $2$D time-reversal symmetry-breaking topological skyrmion phases along the $k_z$ axis. 

First, we consider arguably the simplest non-trivial scenario for values of $\mc{C}$ and $\mc{Q}$, which is $\mc{Q} = -\mc{C}/2$\cite{liu2023}. We show change in $\mc{Q}$ as a function of $k_z$, corresponding to \textit{skyrmion} Weyl nodes, yields novel topological signatures even in these restricted scenarios where $\mc{C} = -2 \mc{Q}$. Later, we also consider a more general topological semimetal due to changes in $\mc{Q}$ vs. $k_z$, with $\mc{C}=0$ for each value of $k_z$, to further illustrate the potential of this non-trivial topology in realizing novel phenomena.

\begin{figure}[t!]
\centering
\includegraphics[width=0.5\textwidth]{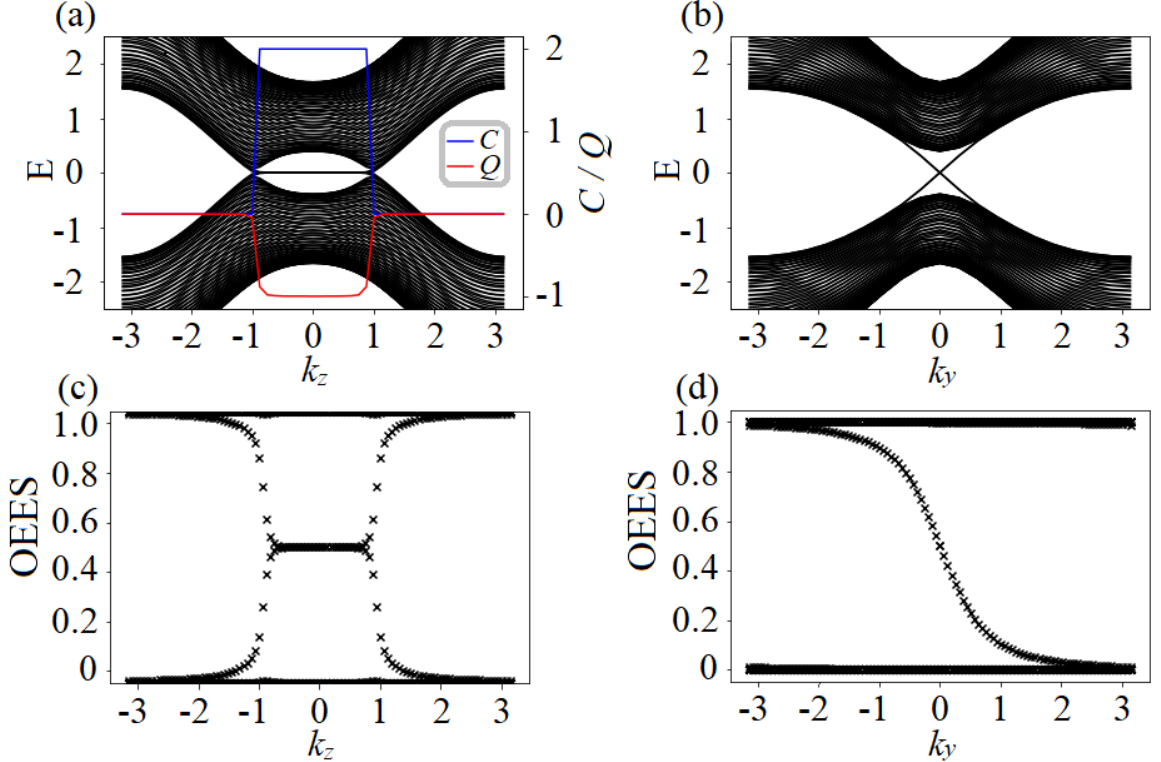}
\caption{(a) Slab energy spectrum of the four-band skyrmion semimetal Hamiltonian Eq.~\ref{eqn: minimal 4-band Weyl} as a function of $k_z$ with fixed $k_y=0$ and open boundary conditions and $N_x = 40$ layers in the $\hat{x}$-direction. We set $\gamma=0.54$ in all numerical calculations, such that the skyrmion number $\mc{Q}$ changes in value at $k_z = \bk^*_\pm=\pm \cos^{-1}(\gamma)=\pm 1$ where gapless modes appear with open boundary conditions. Topological phase transition at the skyrmion nodes as indicated by the Chern number ($C$, blue line) and the skyrmion number ($Q$, red line). (b) Slab energy spectrum with fixed $k_z=0$ as a function of $k_y$ for open-boundary conditions and $N_x = 40$ layers in the $\hat{x}$-direction. Observable-enriched entanglement spectra of the skyrmion semimetal are shown in (c) with fixed $k_y=0$ and (d) at fixed $k_z=0$.} 
\label{fig: ES+OEES}
\end{figure}
We first characterize bulk-boundary correspondence of the TSS Hamiltonian by computing the slab energy spectrum shown in Fig.~\ref{fig: ES+OEES}(a).  We find gapless surface states for the interval of $k_z$ with non-trivial $\mc{C}$ and $\mc{Q}$, similar to the case of a Weyl semimetal. For fixed $k_z$ in this interval, we also show the slab energy spectrum vs. $k_y$ for OBC in the $\hat{x}$-direction, which show $\mc{C}$ chiral states localized on each edge in Fig.~\ref{fig: ES+OEES}(b), similarly to Fermi arc surface states of WSMs. 

\textit{Observable-enriched entanglement spectrum \textemdash} For the topological skyrmion semimetal Hamiltonian Eq.~\ref{eqn: minimal 4-band Weyl}, however, it is possible to further characterize bulk-boundary correspondence and reveal consequences of $\mc{Q}$ even in this very restricted case. We first apply methods of observable-enriched entanglement (OEE) introduced in  Winter~\emph{et al.}~\cite{winter2023}, performing a virtual cut over real-space as in the case of the standard entanglement spectrum~\cite{li2008, alexandradinata2011, zhou2023, kitaev2006a, levin2006, hamma2005, Flammia2009, Thomale2010, Thomale2010b, pollmann2010, turner2010, prodan2010, hughes2011, regnault2009, kargarian2010, Lauchli2010a, Lauchli2010b, bergholtz2011, sterdyniak2011, Rodriguez2010, papic2011, Chandran2011, papic2009, qi2012, zhao2011, schliemann2011, thomale2011, poilblanc2010, turner2012, fidkowski2010b, yao2010, pollmann2010b, calebrese2008, fagoti2011, stephan2011, poilblanc2011, franchini2010, huang2011, cirac2011, dubail2011, liu2011, deng2011, ryu2006}, as well as a virtual cut between degrees of freedom (dofs). The second cut is a modification of the standard partial trace over degrees of freedom which are not spin, which is determined by the spin representation, hence `observable-enriched'. This method of observable-enriched partial trace is reviewed in the Supplementary Materials, Section 1: Observable enriched auxiliary system and entanglement spectrum.

We show OEES vs. $k_z$ for $k_y=0$ and OEES vs. $k_y$ for $k_z=0$ in Figs.~\ref{fig: ES+OEES}(c) and (d), respectively, for direct comparison with Figs.~\ref{fig: ES+OEES} (a) and (b), tracing out half of the system in real-space as well as the generalized particle-hole dof. In Fig.~\ref{fig: ES+OEES}(c), the merging of the top states (OEES=1) and bottom states (OEES=0) at $k_\pm^*=\pm 1$ corresponds to formation of Fermi-arc-like states in the OEES. In Fig.~\ref{fig: ES+OEES}(d), we show that there are also $\mc{Q}$ chiral modes per edge in the OEES\cite{alexandradinata2011} over the interval in $k_z$ for which the OEES exhibits Fermi-arc-like states. These OEES signatures indicate that the spin dof of the four-band model itself \textit{realizes a topological semimetal phase}, specifically due to non-trivial $\mc{Q}$, \textit{with its own} Fermi arc-like surface states resulting from a separate \textit{spin-specific bulk-boundary correspondence}. The Fermi arc surface states in the full four-band model are in fact required in order to yield this bulk-boundary correspondence of the spin subsystem.

\textit{Chiral anomaly of spin degree of freedom\textemdash}We now study response signatures of the TSS when subjected to an external magnetic field $\textbf{B} = (0,0,B)$, to investigate whether the TSS realizes signatures analogous to the chiral anomaly~\cite{Jia2016, pal2023}. The eigenvalues for the two lowest Landau levels (LLLs) can be analytically calculated as detailed in the Supplementary Materials, Section 2: Analytic calculation of Landau levels of topological skyrmion semimetal, and the results are
\begin{equation}
    E_\pm(k_z) = \pm\left( \gamma - \cos{k_z} + \frac{eB}{2} \right).
    \label{eqn: LL}
\end{equation}
\begin{figure}[t!]
    \centering
    \includegraphics[width=0.5\textwidth]{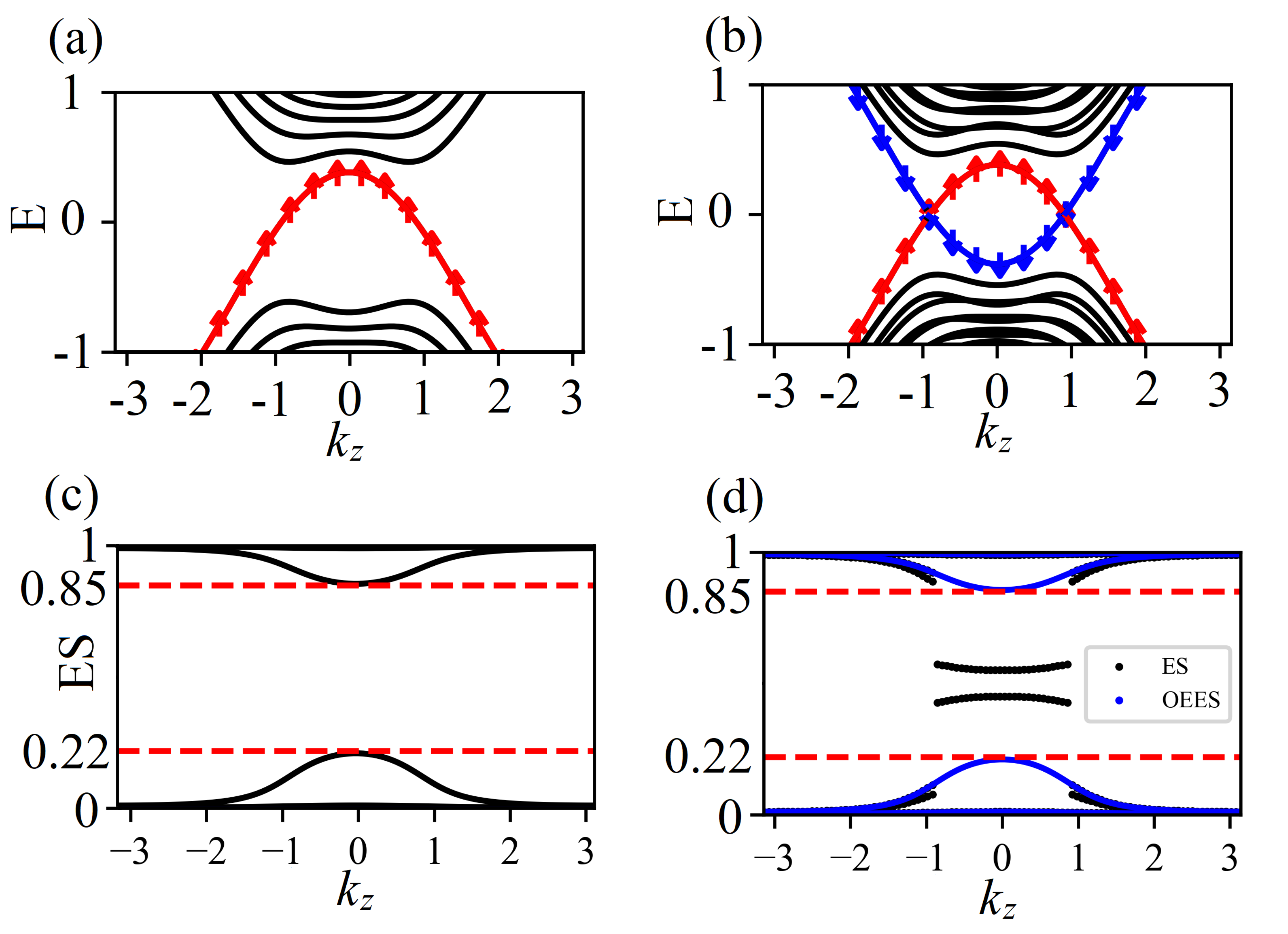}
    \caption{ (a) The Landau levels of the two-band Weyl semimetal phase $h(\bk)$ in Eq.~\eqref{eqn: minimal 4-band Weyl} obtained by numerical diagonalization (solid lines) where the red line is the lowest Landau level, and analytical solution $E_+$ in Eq.~\eqref{eqn: LL}(red up-arrows). (b) The Landau levels of the 4-band skyrmion semimetal Eq.~\eqref{eqn: minimal 4-band Weyl} obtained by numerical diagonalization (solids lines) and analytical solution $E_\pm$ Eq.~\eqref{eqn: LL}(red up-arrows and blue down-arrows respectively). The lowest Landau levels always have spin-texture aligned in the $+z$-axis (up-arrows) or $-z$-axis (down-arrows). (c) The entanglement spectrum of the two-band Weyl semimetal with applied B field where the asymmetry in the band structure manifests as the difference in the magnitude (red dashed lines) of the non-trivial states. (d) The entanglement spectrum (black lines) and observable enriched entanglement spectrum (blue lines) of the 4-band skyrmion semimetal with applied B field where the symmetry in the band structure manifests as the equal magnitude but the OEES remains asymmetric and retains the chiral anomaly signature.}
    \label{fig: LL}
\end{figure}
We also compute the full Landau level (LL) spectrum numerically and compare this with the analytical expressions for LLLs in Fig.~\ref{fig: LL}(a) for the two-band WSM ($E_+$) and in Fig.~\ref{fig: LL}(b) for the four-band TSS ($E_\pm$), respectively. In the latter case, the two LLLs (red and blue) form a generalized charge conjugate pair, so we compute the ES for the WSM and the TSS, as well as the OEES for the TSS, to further probe how these LLLs might combine under observable-enriched partial trace over the generalized particle-hole dof. The ES of the WSM subjected to external magnetic field is shown in Fig.~\ref{fig: LL}(c), which shows the chiral anomaly corresponds to an asymmetry in the ES across the value $0.5$. The ES and OEES of the TSS are shown in Fig.~\ref{fig: LL}(d) for comparison. While the ES is symmetric about the value $0.5$, the OEES is asymmetric similarly to the ES of the WSM, indicating the presence of a chiral anomaly for the spin subsystem due to the TSS phase.

\begin{figure}[t!]
    \centering
    \includegraphics[width=0.5\textwidth]{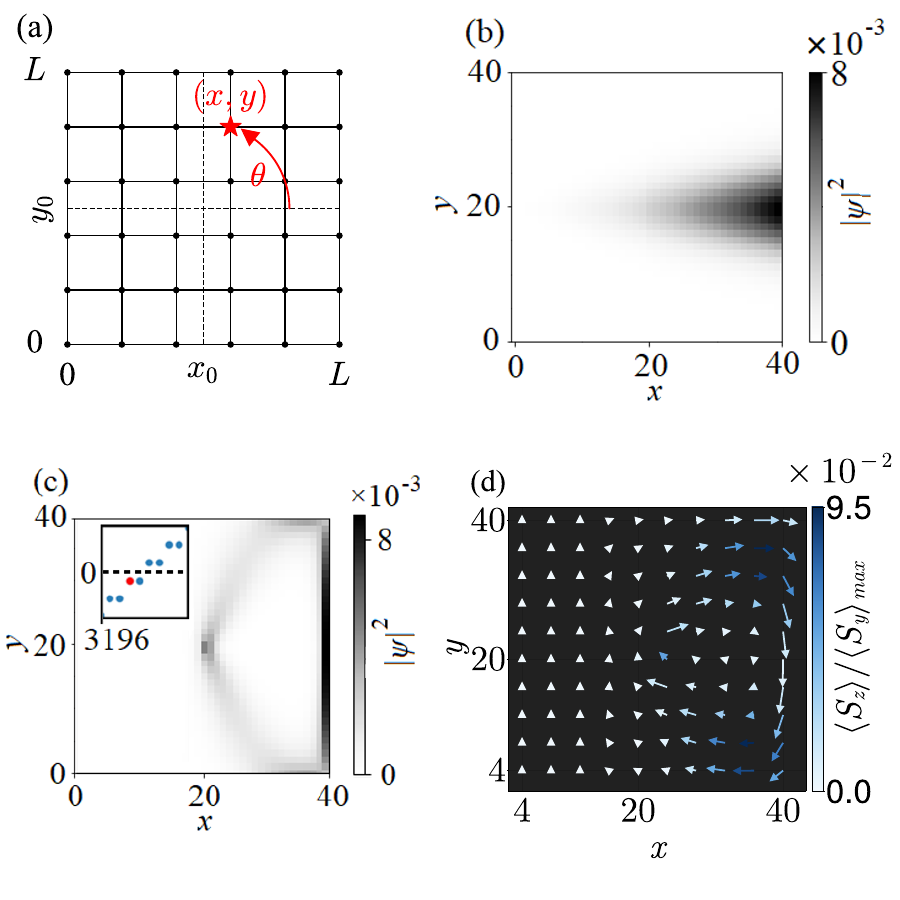}
    \caption{(a) The two-dimensional lattice with the defect located at $(x_0,y_0)$ and $\theta=\arctan[(y-y_0)/(x-x_0)]$. (b) Probability density of the approximate solution of the edge state $|\psi(x,y)|^2$ for small $\gamma$. Numerical results for $\gamma=0.54$ are shown in (c) for probability density vs. $x$ and $y$ for the eigenstate highlighted in red in the inset, which shows spectrum for OBCs vs. eigenvalue index, and in (d) for $\langle \boldsymbol{S} (x,y) \rangle$ of this eigenstate. In (d), the spin magnitude is rescaled by $\langle S_y \rangle_{\textnormal{max}}$, which is the largest $\langle S_y \rangle$ over real space and the larger of $\langle S_x \rangle_{\textnormal{max}}$ and $\langle S_y \rangle_{\textnormal{max}}$, so that all the in-plane spin components $\langle S_x(x,y) \rangle$ and $\langle S_y(x,y)\rangle$ will have length less than $1$ and have no vector overlap.}
    \label{fig:geometry}
\end{figure}

\textit{Bulk-boundary correspondence \textemdash}We now explore the bulk-boundary correspondence of the skyrmion semimetal for a second set of open boundary conditions considered in previous studies of the Hopf insulator~\cite{yan2017} and 3D chiral topological skyrmion phase\cite{liu2023}, but not for topological semimetals, to our knowledge. In this case, we open boundary conditions in the $\hat{x}$- and $\hat{y}$-directions, while retaining periodic boundary conditions in the $\hat{z}$-direction. We then substitute a spatially-varying angle $\theta (x,y)$ for $k_z$, which can be interpreted as an angle in the $x-y$ plane that characterizes a zero-dimensional defect. These open-boundary conditions are depicted in Fig.~\ref{fig:geometry}(a).

%\textcolor{blue}{We consider such boundary conditions for a number of reasons. First, they have already provided great insight on the role of topological classification of defects~\cite{teo2010} in the context of the topological phases of the ten-fold way~\cite{schnyder2008,Ryu_2010}: 3D topological phases of the ten-fold way can inherently trap 0D defects in the BZ that also play a role in the topology~\cite{yan2017}. Second, such boundary conditions have previously been used to great effect to study bulk-boundary correspondence of topological skyrmion phases~\cite{liu2023}. Third, such boundary conditions have not been considered previously for any topological semimetals, to our knowledge.}

We first develop an effective low-energy theory to investigate bulk-boundary correspondence for these OBCs applied to the two-band Weyl semimetal Hamiltonian $h(\boldsymbol{k})$ for $\gamma$ in the vicinity of zero. The details of this calculation are included in the Supplementary Materials, Section 3: Low-energy theory of Weyl semimetal for 2D system plus defect. The effective Hamiltonian is calculated to be
\begin{align}
   H_{\textnormal{eff}} &= 
   \begin{pmatrix}
    k_x & \frac{1}{2}\theta^2\\
    \frac{1}{2}\theta^2 & -k_x
    \end{pmatrix},
\end{align} 
where $k_x$ is the remaining momentum component and $\theta$ is the angle parameterizing the defect as shown in Fig.~\ref{fig:geometry}(a). 

The approximate wave function we obtain for this finite square lattice with a defect is
\begin{equation}
    |\psi(x,y)\rangle = C \frac{x e^{-y^2/2x}}{y^2}
\end{equation}
where $C$ is the normalization constant. The probability density is shown in Fig.~\ref{fig:geometry}(b) and is a good approximation of the numerical results shown in Fig. \ref{fig:geometry} c). In Fig.~\ref{fig:geometry}(c) we show the probability density of one of the lowest energy states (i.e. $|E|$ closest to 0) with $\gamma=0.54$ for the full tight-binding Hamiltonian. Probability density for this state peaks along the right edge, but also extends along portions of the top and bottom edges up to  $\theta=\bk_\pm^*\approx \pm57$\textdegree, before finally leaking into the bulk at these values of $\theta$ as they correspond to the positions of the gapless points in the bulk spectrum. In Fig.~\ref{fig:geometry}(d), (e), and (f), we show the three components of the spin texture for the same state considered in Fig.~\ref{fig:geometry} (c). The two-band Weyl semimetal Hamiltonian displays similar bulk-boundary correspondence with these boundary conditions. We expect the edge states of the TSS Hamiltonian to form generalized charge conjugate pairs which combine under observable-enriched partial trace to yield edge states for the spin subsystem similar to those of the Weyl semimetal, but degenerate states have spin textures with the same structure in the $z$-component but opposite sign for $x$- and $y$-components, such that there is naively an ambiguity in the outcome of tracing out the generalized particle-hole dof. One can break the  degeneracy of the zero-energy manifold by introducing a magnetic field in the $\hat{z}$-direction along the edge at $x=L$, however. The resultant energy levels and spin textures are provided in the Supplementary Materials, Section 4: Spin texture of edge states in defect square lattice, and we see the generalized charge conjugate pairs indeed combine to yield a generalized Weyl semimetal phase of the spin subsystem. 
\begin{figure}[t]
    \centering
    \includegraphics[width=0.5\textwidth]{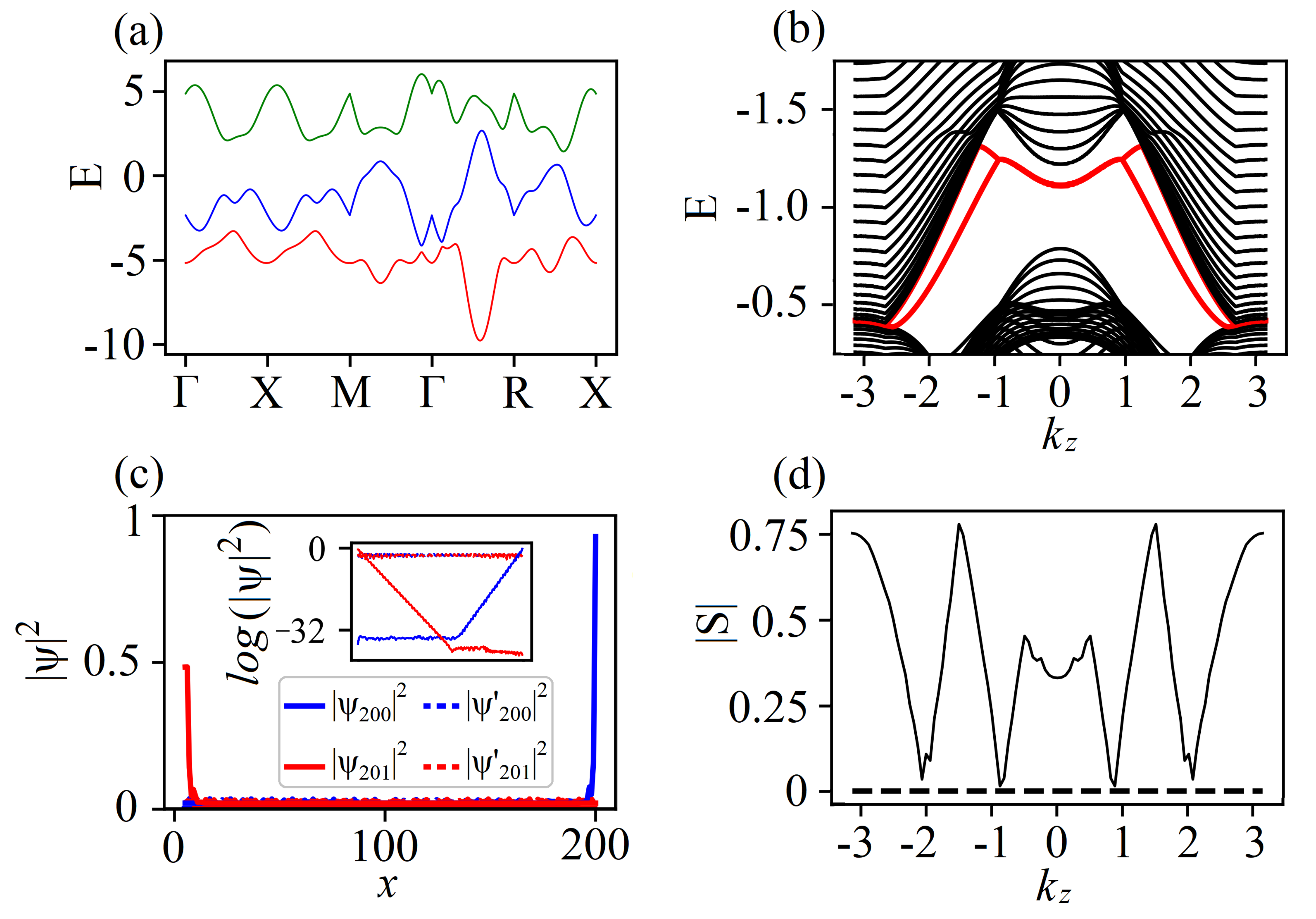}
    \caption{(a) Bulk energy spectrum plotted against a linearly interpolated trajectory along high-symmetry points $\Gamma(0,0,0)$, X$(0,\pi,0)$, M$(\pi,\pi,0)$ and R$(\pi,\pi,\pi)$. (b) Slab energy spectrum for fixed $k_y$ sector with open boundary conditions and $N_x = 40$ unit cells in the $\hat{x}$-direction plotted against $k_z$. We show the $k_y$ sector for each value of $k_z$, which corresponds to the minimum difference in energy between the in-gap states highlighted in red. (c) Probability density vs. layer index $x$ of the eigenstates $\psi_i$  ($\psi'_i$), where $i$ indicates the state corresponds to the $i$\textsuperscript{th} lowest energy eigenvalue, at $k_y=-3\pi/4$ and $k_z=0$ ($k_z=\pi$). Log of probability density vs. $x$ for each state is also shown in the inset. (d) Minimum spin expectation value magnitude over the $k_x$-$k_y$ BZ submanifold for fixed $k_z$ plotted vs. $k_z$.}
    \label{fig: 3band}
\end{figure}

\textit{Three-band skyrmion semimetals \textemdash} We finally construct Hamiltonians for topological skyrmion semimetals from lower-symmetry three-band models for 2D topological skyrmion phases~\cite{cook2023, cook2023QSkHE}. The three-band Bloch Hamiltonian with basis $\Psi_{\boldsymbol{k}} = \left( c_{\boldsymbol{k}, xy, \uparrow}, c_{\boldsymbol{k}, yz, \downarrow}, c_{\boldsymbol{k}, xz, \downarrow} \right)^{\top}$, where $\{xy, yz, xz \}$ label a three-fold $t_{2g}$ orbital dof and $\{\uparrow, \downarrow \}$ label a spin $1/2$ dof, is compactly written as
\begin{equation}
    \mathcal{H}(\bk) = \boldsymbol{d}_1(\bk) \cdot \boldsymbol{\sigma}_1 + \boldsymbol{d}_2(\bk) \cdot \boldsymbol{\sigma}_2 + \lambda \boldsymbol{\sigma}_{3,x}
    \label{3band Ham},
\end{equation}
where $\boldsymbol{\sigma}_{1,2,3}$ are three different embeddings of the Pauli matrix vector into $3\times 3$ matrix representations, $\mathbf{d_1}$ and $\mathbf{d_2}$ are two distinct modifications of the QWZ $\mathbf{d}$-vector for a two-band Chern insulator\cite{qi2006}, and $\lambda$ is a constant. Additional details on the Hamiltonian and spin representation shown in related work introducing the quantum skyrmion Hall effect~\cite{cook2023, cook2023QSkHE} are also provided in the Supplementary Materials, Section 5: Details of three-band Bloch Hamiltonian for 2D chiral topological skyrmion phase.

 In Fig.~\ref{fig: 3band}(a) we show the  bulk energy spectrum along a high-symmetry path through the BZ, which indicates finite minimum direct bulk energy gap between the lowest and second-lowest energy bands and the absence of topological band-touchings. $k_z$ dependence is chosen, however, to yield topological phase transitions according to skyrmion number $\mc{Q}$ at $k_z \approx \pm \pi/3$, while the total Chern number is zero for all values of $k_z$.  Fig.~\ref{fig: 3band}(b) shows a slab energy spectrum for the system with open boundary conditions in the $\hat{x}$- direction, which exhibits in-gap states highlighted in red: we show the spectrum for fixed $k_y$ sector for each value of $k_z$, which corresponds to the minimum difference in energy between the in-gap states highlighted in red. The in-gap states correspond to surface bands crossing in the slab spectrum for fixed $k_z$, yielding gaplessness for $k_z \in \{-1,1 \}$, approximately, in the sense that the Fermi level will always intersect the edge bands while in the bulk energy gap. Gaplessness is lost outside this interval, where the edge states at fixed $k_z$ no longer cross, and the states may then be smoothly deformed into the bulk as shown in Fig.~\ref{fig: 3band}(c).  Importantly, the surface states do not terminate in closure of the bulk energy gap in the form of topological band structure degeneracies, as in the case of a WSM.  In Fig.~\ref{fig: 3band}(d), we demonstrate that the generalized Fermi arc states terminate at type-II topological phase transitions~\cite{cook2023}, in which $\mc{Q}$ changes due to spin becoming zero in magnitude, without closing of the minimum direct bulk energy gap.

\textit{Discussion and conclusion \textemdash }We introduce topological skyrmion semimetal (TSS) phases of matter by constructing toy models in which specifically the spin degree of freedom (dof) in systems with multiple dofs (in this case, a generalized particle-hole dof or orbital dof) can realize generalized Fermi arc surface states and chiral anomaly signatures of the spin subsystem. Remarkably, we utilize three-band models for 2D topological skyrmion phases to construct topological skyrmion semimetal Hamiltonians possessing Fermi arc-like surface states in \textit{bulk insulators}, due to topological defects of the momentum-space spin texture. Our work therefore introduces a fundamental generalization of topological semimetals \textit{and insulators} by considering topological phases associated with mappings to myriad observables, rather than the projectors onto occupied states. The three-band TSS is a very low-symmetry topological state, similar to the Weyl semimetal, which makes it a promising platform for experiments. 

\textit{Acknowledgements \textemdash} We gratefully acknowledge helpful discussions with A.~Pal, R.~Calderon and R.~Ay. This research was supported in part by the National Science Foundation under Grants No. NSF PHY-1748958 and PHY-2309135, and undertaken in part at Aspen Center for Physics, which is supported by National Science Foundation grant PHY-2210452.

\bibliographystyle{apsrev4-2}

\bibliography{p1bib.bib}

\cleardoublepage

%%%%%%%%% Prefix a "S" to all equations, figures, tables and reset the counter %%%%%%%%%%
\makeatletter
\renewcommand{\theequation}{S\arabic{equation}}
\renewcommand{\thefigure}{S\arabic{figure}}
\renewcommand{\thesection}{S\arabic{section}}
\setcounter{equation}{0}
\setcounter{section}{0}
\onecolumngrid
\begin{center}
  \textbf{\large Supplemental material for ``Topological skyrmion semimetals''}\\[.2cm]
  Shu-Wei Liu$^{1,2}$, Joe H. Winter$^{1,2,3}$ and Ashley M. Cook$^{1,2,*}$\\[.1cm]
  {\itshape ${}^1$Max Planck Institute for Chemical Physics of Solids, Nöthnitzer Strasse 40, 01187 Dresden, Germany\\
  ${}^2$Max Planck Institute for the Physics of Complex Systems, Nöthnitzer Strasse 38, 01187 Dresden, Germany\\
  ${}^3$SUPA, School of Physics and Astronomy, University of St.\ Andrews, North Haugh, St.\ Andrews KY16 9SS, UK\\
  ${}^*$Electronic address: cooka@pks.mpg.de\\}
%(Dated: \today)\\[1cm]
\end{center}

\section{Observable-enriched auxiliary system and entanglement spectrum}
%

%The analysis of non-trivial spin textures can be undertaken by reframing the problem in the language of topological band theory through the employment of the auxiliary spin ground state. This entails creating an effective two-level mixed state which has Pauli matrices expectation values equivalent to the expectation values of the relevant operators in the Bloch ground state. In general this can be thought of as moving to an effective measurement basis of the magnetization \cite{winter2023}:

We briefly summarize methods introduced in Winter~\emph{et al.}~\cite{winter2023}, which are employed here to characterize entanglement properties of topological skyrmion semimetal phases of matter. We first consider two-band Bloch Hamiltonians with (pseudo)spin $1/2$ degree of freedom, $\mc{H}(\boldsymbol{k})$,  compactly written as
\begin{equation}
    \mc{H}(\boldsymbol{k}) = \boldsymbol{d}(\boldsymbol{k})\cdot \boldsymbol{\sigma},
\end{equation}
where $\boldsymbol{d}(\boldsymbol{k})$ is a three-vector of momentum-dependent functions $\boldsymbol{d}(\boldsymbol{k}) = \langle d_x(\boldsymbol{k}), d_y(\boldsymbol{k}), d_z(\boldsymbol{k}) \rangle$ and $\boldsymbol{\sigma} = \langle \sigma_x, \sigma_y, \sigma_z \rangle$ is the vector of Pauli matrices. Here, $\{\sigma_i\}$ are also an effective (pseudo)spin representation sufficient to compute (pseudo)spin skyrmion number $\mc{Q}$ in terms of the ground state (pseudo)spin expectation value given by $\boldsymbol{d}(\boldsymbol{k})$ as
\begin{equation}
    \mc{Q} = {1 \over 4\pi} \int d\boldsymbol{k} \left[\boldsymbol{\hat{d}}(\boldsymbol{k}) \cdot \left(\partial_{k_x} \boldsymbol{\hat{d}}(\boldsymbol{k}) \times \partial_{k_y} \boldsymbol{\hat{d}}(\boldsymbol{k}) \right) \right] ,
\end{equation}

for $\boldsymbol{k} = (k_x, k_y)$ defining a two-dimensional Brillouin zone and $\boldsymbol{\hat{d}}(\boldsymbol{k}) = \boldsymbol{d}(\boldsymbol{k})/|\boldsymbol{d}(\boldsymbol{k})|$ the normalized ground-state spin expectation value of the two-band Bloch Hamiltonian $\mc{H}(\boldsymbol{k})$. In this case, $\mc{Q} = \mc{C}$, the Chern number of the lower-energy band, and there are $\mc{Q}$ chiral modes localized on the boundary in correspondence. Given this, the $\boldsymbol{d}(\boldsymbol{k})$-vector also defines a density matrix in each $\boldsymbol{k}$-sector, $\rho(\boldsymbol{k})$, which, for general $N$-band systems, winds over the Brillouin zone in correspondence with the total Chern number.

For the four-band Bloch Hamiltonians with generalized particle-hole  symmetry $\mc{C}'$ and spin $1/2$ degree of freedom discussed in the present work, $\mc{Q}$ and $\mc{C}$ can be independent topological invariants, with $\mc{Q}$ still computed as the skyrmion number in terms of the ground-state spin expectation value $\langle \boldsymbol{S}\rangle = \left(\langle S_x\rangle, \langle S_y\rangle, \langle S_z\rangle \right)$ defined over the Brillouin zone as stated in Eq. 4 in the main text. In analogy to the topological characterization of the two-band Hamiltonian $\mc{H}(\boldsymbol{k})$ in terms of density matrix $\rho(\boldsymbol{k})$, we may then define an effective bulk two-level system in terms of an auxiliary density matrix $\rho^S(\boldsymbol{k})$ computed from the spin expectation value of the four-band system in each $\boldsymbol{k}$-sector as
\begin{equation}
    \rho^{S}(\boldsymbol{k}) = \frac{1}{2} ( \mathbf{I}_2 + \text{Tr}\left[ \rho(\boldsymbol{k}) \boldsymbol{S} \right] \cdot \boldsymbol{\sigma} ), \label{eq:OEGS}
\end{equation}
where $\mathbf{I}_2$ is the $2 \times 2$ identity matrix.

%\textcolor{red}{In systems which the spin is separable between the underlying Hilbert spaces of the degrees of freedom we can simplify this mapping to a partial trace. In essence, systems where spin is of the form $S_{\mu} = \mathbb{I}_{\tau} \otimes \sigma_{\mu}$ and the Hilbert space is formed of subspaces $\mathcal{H}_{\tau}, \mathcal{H}_{\sigma}$. In this case, we calculate the auxiliary system with a partial trace over the $\tau$ subspace. For the system outlined in Eq.~2 the spins, defined by $S^{\mu} = \text{diag}(\sigma_{\mu}, -\sigma_{\mu}^*)$, are isomorphic to $\mathbb{I}_2 \otimes \sigma_{\mu}$ by change of basis $\mathbb{I} \oplus \sigma_y$. Therefore, performing this change of basis and partial tracing the ground state projector over the new non spin degrees of freedom we may produce the auxiliary spin ground state defined by Eq.~\ref{eq:OEGS}.}\par 

%Furthermore, the use of reduced density matrices allows analysis over entanglement cuts in real space. Here we expect spectral flow between the entanglement ground and entanglement excitation spectra. This is due to the inability to localise the auxiliary entanglement wavefunctions which produce the underlying topological spin texture \cite{winter2023,alexandradinata2011}. 
More broadly, we may define $\rho^{S}$ as an effective reduced density matrix derived  from $\rho$ of the full system directly in terms of a generalized partial trace operation. That is, rather than performing a partial trace in general, computation of $\rho^S$ directly from $\rho$ is enriched by the spin representation of the full system as detailed in Winter~\emph{et al.}~\cite{winter2023}

To characterize entanglement of the topological skyrmion semimetals, we define an auxiliary spin ground state density matrix for a slab geometry (open boundary conditions in the $\hat{x}$-direction and periodic boundary conditions in the $\hat{y}$-direction) via Fourier transform as
\begin{equation}
    \rho^S_{\text{slab}} = \sum_{x,x'} \frac{1}{2} (\mathbf{I}_{2} + \text{Tr}[\rho_{x,x'} \boldsymbol{S}] \cdot \boldsymbol{\sigma}) \ket{x}\bra{x'},
\end{equation}
where $\rho_{x,x'}$ are the matrix elements $\braket{x|\rho_{\text{slab}}|x'}$ computed from the density matrix of the full four-band system in this slab geometry, $\rho_{\text{slab}}$, with real-space layer indices $x,x'$. Entanglement spectra are then produced from $\rho^{S}_{\text{slab}}$ via the method of Peschel~\cite{peschel2009} as in past work characterizing band topology~\cite{alexandradinata2011}.

\section{Analytic calculation of Laudau levels of topological skyrmion semimetal}\label{Sup_sec_I}
We analytically calculate the Landau levels in the case of no spin triplet pairing term, so that the diagonal blocks $h(\bk)$ and $-h^*(\bk)$ can be solved separately. Take the expansion of $h(\bk)$ around $\bk=0$:
\begin{equation}
    h(\bk) = k_x \sigma_x + k_y \sigma_y +\left[ \gamma - \cos(k_z)+\frac{1}{2} \left( k_x^2 + k_y^2 \right) \right] \sigma_z.
\end{equation}
Now we take the following gauge transformation:
\begin{align*}
k_x \rightarrow k_x, \hspace{0.25cm}
k_y \rightarrow k_y + eBx, \hspace{0.25cm}
k_z \rightarrow k_z,
\end{align*}
so that we construct lowering and raising operators $a=\frac{k_y-ik_x}{\sqrt{2eB}}$ and $a^\dagger=\frac{k_y+ik_x}{\sqrt{2eB}}$ such that the following commutation relation is satisfied
\begin{equation}
    [a,a^\dagger] = 1,
\end{equation}
and $h(\bk)$ can be recast to 
\begin{align}
    h(\bk) = &\sqrt{2eB}\left( a\sigma_+ + a^\dagger\sigma_- \right) \nonumber\\
    &+\left[ \gamma - \cos(k_z) + eB \left( a^\dagger a + \frac{1}{2} \right) \right] \sigma_z,
    \label{eqn: operator}
\end{align}
where
\begin{equation*}
    \sigma_\pm = \frac{\sigma_x \pm i \sigma_y}{2}
\end{equation*}
and
\begin{align*}
    \sigma_z |\pm \rangle = \pm|\pm \rangle, \hspace{0.3cm}
    \sigma_\pm |\mp \rangle = |\pm \rangle.
\end{align*}
We can therefore express the lowest Landau level as $|0,-\rangle$ where the first index denotes the energy level and the second index denotes the spin. The form of Eq.~\eqref{eqn: operator} is useful because the $a$ operator in the $a\sigma_+$ term acts on and annihilates the energy part of the ground state the $\sigma_-$ operator in the $a^\dagger\sigma_-$ term annihilates the spin part of the ground state. The ground state energy is therefore conveniently
\begin{equation}
    E(k_z) = -\left( \gamma - \cos{k_z} + \frac{eB}{2} \right).
    \label{eqn: 2-band LL}
\end{equation}
As for $-h^*(\bk)$, the only difference is that the $\sigma_x$ and $\sigma_z$ terms pick up a minus sign, so we have 
\begin{align}
    -h^*(\bk) = &\sqrt{2eB}\left( a^\dagger\sigma_+ + a\sigma_- \right) \nonumber\\
    &-\left[ \gamma - \cos(k_z) + eB \left( a^\dagger a + \frac{1}{2} \right) \right] \sigma_z,
\end{align}
with the lowest Landau level $|0,+\rangle$ and the energy
\begin{equation}
    E(k_z) = +\left( \gamma - \cos{k_z} + \frac{eB}{2} \right)
    \label{4-band LL}
\end{equation}

\section{Low-energy theory of Weyl semimetal for 2D system plus defect}\label{Sup_sec_II}
Here we derive a low-energy theory for the two-band Weyl semimetal phase with Hamiltonian defined in Eq.(1), for the specified boundary condition in this paper and a semi-infinite geometry. In order to approximate the wave functions around $x=0$, translational symmetry in the $\hat{y}$-direction is therefore broken while momentum component $k_x$ remains as a good quantum number. Thus, a real-space coordinate is used in the $\hat{y}$-direction to label lattice sites. In addition, the specified geometry requires that the topological phase transition takes place at $\pm \bk^*=\pm \cos^{-1}{\gamma}$, so to keep zero-energy states close to $x=0$, we denote $\gamma=1-\Delta \gamma$ where $\Delta \gamma$ is a small quantity. The wave functions and energy eigenvalues can be obtained by solving the following equation
\begin{equation}
    \begin{pmatrix} 
    H_0 & H_1 & 0 & 0 & \cdots \\
    H_1^\dagger & H_0 & H_1 & 0 & \cdots \\
    0 & H_1^\dagger & H_0 & H_1 & \cdots \\
     \vdots & \vdots & \vdots & \vdots & \ddots \\
    \end{pmatrix} 
    \begin{pmatrix}
    \psi_1 \\
    \psi_2 \\
    \psi_3 \\
    \vdots
    \end{pmatrix}
    =
    E
    \begin{pmatrix}
    \psi_1 \\
    \psi_2 \\
    \psi_3 \\
    \vdots
    \end{pmatrix}
    \label{eqn:main}
\end{equation}
where $\psi_j$ is the $j^{th}$ entry of an eigenstate for this equation corresponding to the $j^{th}$ lattice site in the $\hat{x}$-direction. Each $\psi_j$ has two components corresponding to the spin degrees of freedom at each site in real-space, and the non-zero terms of the Hamiltonian matrix representation on the left-hand side of Eq.~\ref{eqn:main} take the following forms: the term on the diagonal is

\begin{equation}
    H_0(k_y,\lambda)
    =
    \begin{pmatrix}
    2-\cos{k_y}+1-\Delta \gamma-\cos{\lambda} & \sin{k_x}\\
    \sin{k_x} & -(2-\cos{k_y}+1-\Delta \gamma-\cos{\lambda})\\
    \end{pmatrix},
\end{equation}
and term off the diagonal is

\begin{equation}
     H_1
    =
    -\frac{1}{2}\begin{pmatrix}
    1 & 1\\
    -1 & -1\\
    \end{pmatrix},
\end{equation}
where $\lambda$ is a parameter substituted for the $\hat{z}$-component of the momentum, $k_z$, characterizing a defect. Here, $\lambda=n\theta(i,j)$ with $\theta$ being the relative angle between the defect and site$(i,j)$, $n$ is an integer taken to be $1$ in this study.

To find zero-energy solutions in analogy to Yan~\emph{et al.}~\cite{yanzhongbo2017}, we first set $k_x=0$ and $\lambda=0$. The non-zero entries of the Hamiltonian in Eq.~\ref{eqn:main} then take the following forms:
\begin{equation}
    H_0(0,0)
    =
    (1-\Delta \gamma) 
    \begin{pmatrix}
    1 & 0\\
    0 & -1
    \end{pmatrix}\\
\end{equation}

We then consider eigenstates of $\sigma_x$, which are
\begin{equation}
\begin{aligned}
    |\nu_1\rangle &= \frac{1}{\sqrt{2}} \begin{pmatrix}
    1 \\
    1 
    \end{pmatrix}\\
    |\nu_2\rangle &= \frac{1}{\sqrt{2}} \begin{pmatrix}
    1 \\
    -1 
    \end{pmatrix}
\end{aligned}    
\end{equation}
with the eigenvalues $\lambda_{1}=+1$ and $\lambda_{2}=-1$. Operating the matrices $H_0, H_1$ and $H_1^\dagger$ on $|\nu_1\rangle$ and $|\nu_2\rangle$, they produce:
\begin{equation}
\begin{aligned}
    H_0 
    \begin{pmatrix}
    |\nu_1\rangle \\
    |\nu_2\rangle \\
    \end{pmatrix}
    &=
    (1-\Delta \gamma)
    \begin{pmatrix}
    -|\nu_2\rangle \\
    |\nu_1\rangle \\
    \end{pmatrix}\\
    H_1 
    \begin{pmatrix}
    |\nu_1\rangle \\
    |\nu_2\rangle \\
    \end{pmatrix}
    &=
    \begin{pmatrix}
    -|\nu_2\rangle \\
    0 \\
    \end{pmatrix}\\
    H_1^\dagger
    \begin{pmatrix}
    |\nu_1\rangle \\
    |\nu_2\rangle \\
    \end{pmatrix}
    &=
    \begin{pmatrix}
    0 \\
    -|\nu_1\rangle \\
    \end{pmatrix}
\end{aligned}
\end{equation}
The zero-energy wave functions take the form
\begin{equation}
    |\Psi_i\rangle = \sum_j a_{i,j} |j\rangle \otimes |\nu_i\rangle
\end{equation}
where $a_{i,j}$ is a function of $j$ that normalizes $|\Psi_i\rangle$ and $|j\rangle$ indicates localization on the $j^{th}$ site:
\begin{equation}
    |j\rangle = (0,0,\cdots,0,1,0,\cdots)^\top
\end{equation}
i.e. having $1$ in the $j^{th}$ element and $0$ everywhere else. We can explore different possibilities of $|\Psi_i\rangle$ by substituting this expression in Eq.\eqref{eqn:main} and consider the $j^{th}$ element of the equation.\\
In the case of $i=1$:
\begin{equation}
     H_1^\dagger a_{1,j-1} |\nu_1\rangle + H_0 a_{1,j} |\nu_1\rangle + H_1 a_{1,j+1} |\nu_1\rangle = E = 0
\end{equation}
which simplifies to
\begin{equation}
    \left[ 0+(1-\Delta \gamma)a_{1,j}-a_{1,j+1} \right]|\nu_2\rangle = 0
\end{equation}
and therefore obtaining the following recurrence relation in $a_{1,j}$:
\begin{equation}
    a_{1,j+1} = (1-\Delta \gamma)a_{1,j}
\end{equation}
which solves to give the general formula:
\begin{equation}
    a_{1,j} = (1-\Delta \gamma)^{j-1}a_{1,1}
\end{equation}
where $a_{1,1}$ is effectively the normalization constant. \\
In the case of $i=2$:
\begin{equation}
     H_1^\dagger a_{2,j-1} |\nu_2\rangle + H_0 a_{2,j} |\nu_2\rangle + H_1 a_{2,j+1} |\nu_2\rangle = E = 0
\end{equation}
which simplifies to
\begin{equation}
    \left[-a_{2,j-1}+(1-\Delta \gamma)a_{2,j}+0\right]|\nu_1\rangle = 0
\end{equation}
and therefore obtaining the following recurrence relation in $a_{2,j}$:
\begin{equation}
    a_{2,j+1} = \frac{1}{1-\Delta \gamma} a_{2,j}.
\end{equation}
which solves to give the general formula:
\begin{equation}
    a_{2,j} = \frac{1}{(1-\Delta \gamma)^{j-1}}a_{2,1}
\end{equation}
where $a_{2,1}$ is again effectively the normalization constant. \\
To satisfy the orthonormality condition of the zero-energy wave functions
\begin{equation}
\langle\Psi_i|\Psi_j\rangle = \delta_{ij},    
\end{equation}
it suffices to note that for $i=1,2$, $\sum_{j=1}^\infty (a_{i,j})^2=1$ because $|\nu_1\rangle$ and $|\nu_2\rangle$ are eigenstates of $\sigma_x$ so $\langle\nu_i|\nu_j\rangle = \delta_{ij}$.

Next, we consider the Hamiltonian in the neighbourhood of $k_x=0, \lambda=0$ by expanding $H$ to first order of these variables i.e. $H_{0/1}(k_x,\lambda)=H_{0/1}(0,0)+\Delta H_{0/1}(k_x,\lambda)$ where
\begin{align}
    \Delta H_0 
    &= 
    \begin{pmatrix}
    \frac{1}{2}\lambda^2 & k_x\\
    k_x & -\frac{1}{2}\lambda^2
    \end{pmatrix}\\
    \Delta H_1 
    &= 
    0,
\end{align}
so that
\begin{equation}
    \Delta H = \begin{pmatrix} 
    \Delta H_0  & 0 & 0 & \cdots \\
    0 & \Delta H_0 & 0 & \cdots \\
    0 & 0 & \Delta H_0 & \cdots \\
    \vdots  & \vdots & \vdots & \ddots \\
    \end{pmatrix}. 
\end{equation}
and the low-energy effective Hamiltonian is
\begin{align}
    H_{\textnormal{eff}} 
    &= 
    \begin{pmatrix}
    \langle\Psi_1|\Delta H|\Psi_1\rangle & \langle\Psi_1|\Delta H|\Psi_2\rangle & \\
    \langle\Psi_2|\Delta H|\Psi_1\rangle & \langle\Psi_2|\Delta H|\Psi_2\rangle
    \end{pmatrix}.
\end{align}

The matrix elements are thus calculated as:
\begin{align}
    \langle\Psi_1|\Delta H|\Psi_1\rangle 
    &= \sum_{j=1}^\infty a_{1,j}\frac{1}{\sqrt{2}}
    \begin{pmatrix}
    1\\
    1
    \end{pmatrix} 
    \begin{pmatrix}
    \frac{1}{2}\lambda^2 & k_x\\
    k_x & -\frac{1}{2}\lambda^2
    \end{pmatrix} 
    a_{1,j}\frac{1}{\sqrt{2}}\begin{pmatrix}
    1\\
    1
    \end{pmatrix}\\
    &=k_x \sum_{j=1}^\infty a_{1,j}^2 \nonumber\\
    &=k_x \nonumber\\
    \langle\Psi_1|\Delta H|\Psi_2\rangle 
    &= \sum_{j=1}^\infty a_{1,j}\frac{1}{\sqrt{2}}
    \begin{pmatrix}
    1\\
    1
    \end{pmatrix} 
    \begin{pmatrix}
    \frac{1}{2}\lambda^2 & k_x\\
    k_x & -\frac{1}{2}\lambda^2
    \end{pmatrix} 
    a_{2,j}\frac{1}{\sqrt{2}}\begin{pmatrix}
    1\\
    -1
    \end{pmatrix}\\
    &=\frac{1}{2}\lambda^2\\
    \langle\Psi_2|\Delta H|\Psi_1\rangle
    &= (\langle\Psi_1|\Delta H|\Psi_2\rangle)^*\\
    &= \frac{1}{2}\lambda^2\\
    \langle\Psi_2|\Delta H|\Psi_2\rangle 
    &= \sum_{j=1}^\infty a_{2,j}\frac{1}{\sqrt{2}}
    \begin{pmatrix}
    1\\
    -1
    \end{pmatrix} 
    \begin{pmatrix}
    \frac{1}{2}\lambda^2 & k_x\\
    k_x & -\frac{1}{2}\lambda^2
    \end{pmatrix} 
    a_{2,j}\frac{1}{\sqrt{2}}\begin{pmatrix}
    1\\
    -1
    \end{pmatrix}\\
    &=k_x \sum_{j=1}^\infty a_{2,j}^2 \nonumber\\
    &=k_x \nonumber\\
\end{align}

Therefore the effective Hamiltonian is
\begin{align}
   H_{\textnormal{eff}} &= 
   \begin{pmatrix}
    k_x & \frac{1}{2}\lambda^2\\
    \frac{1}{2}\lambda^2 & -k_x
    \end{pmatrix}
\end{align}
so that
\begin{align}
   H_{\textnormal{eff}}^2 &= 
   \begin{pmatrix}
    k_x^2+\frac{\lambda^4}{4} & 0\\
    0 & k_x^2+\frac{\lambda^4}{4}
    \end{pmatrix}
    = \left( k_x^2+\frac{\lambda^4}{4} \right) \mathbf{I}_2
\end{align}

Squaring both sides of the time-independent Schrödinger equation and taking $H_{\textnormal{eff}} (k_x\rightarrow -i\partial_x, \lambda \rightarrow y/x)$ yields 
\begin{align}
\left( -\partial_x^2 + \frac{y^4}{4x^4} \right) \mathbf{I}_2 |\psi\rangle = E^2 |\psi\rangle,
\end{align}

The eigenstates of this low-energy effective Hamiltonian may then be expressed as $|\mu_i \rangle = \chi_i^{\top} |\psi \rangle$, where $\chi_i$ satisfies the constraints of the Hamiltonian matrix representation, and $|\psi\rangle = |\psi(x,y)\rangle$ is a spatially-varying scalar function satisfying the differential equations contained in the Hamiltonian. We first identify $\chi_1=(1,0)$ and $\chi_2=(0,1)$ are the eigenvectors of $\mathbf{I}_2$.

These states serve as zero-energy eigenstates of the effective low-energy Hamiltonian if the same differential equation is satisfied:
\begin{align}
    \left(-\partial_x^2 + \frac{y^4}{4x^4}\right)|\psi\rangle &= 0
\end{align}
which has the solution
\begin{align}
    |\psi\rangle &= C_1xe^{y^2/2x} + \frac{C_2xe^{-y^2/2x}}{y^2}
\end{align}
where we reject the first term which is unnormalizable due to the exponential term.

\newpage
\section{Spin texture of edge states in defect square lattice}\label{Sup_sec_III}
 Here we present fully the probability density and spin texture of the edge states shown in Fig.~3 in the main text:
\begin{figure}[h!]
    \centering
    \includegraphics[width=1.0\textwidth]{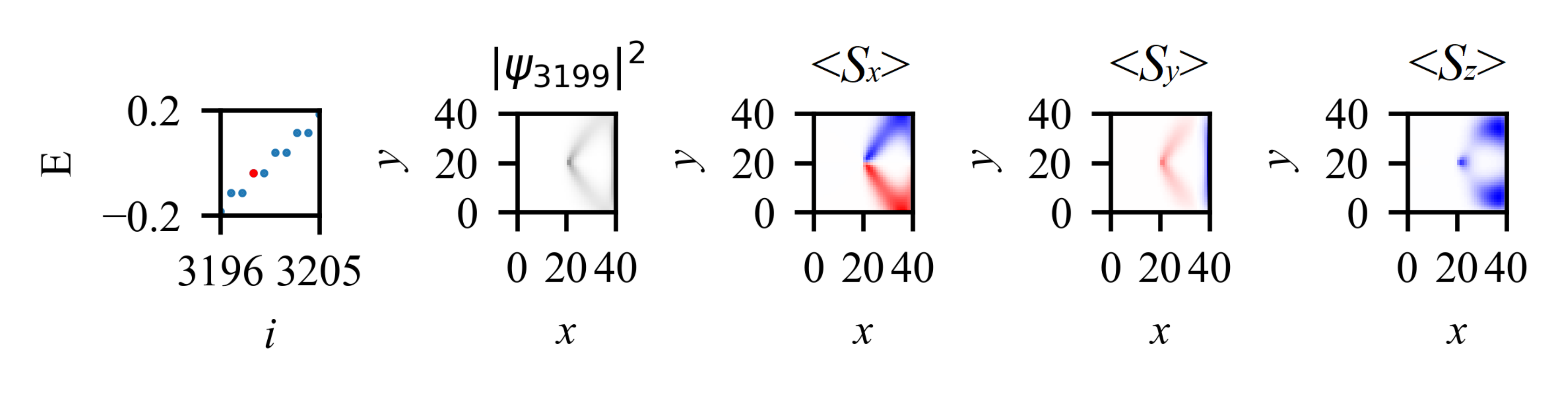}
    \includegraphics[width=1.0\textwidth]{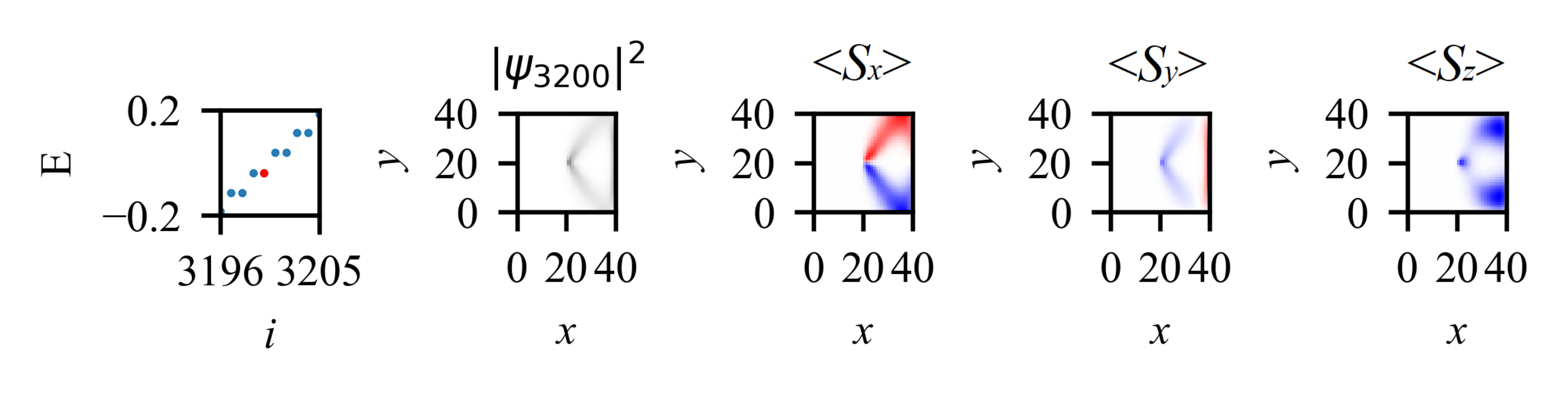}
    \includegraphics[width=1.0\textwidth]{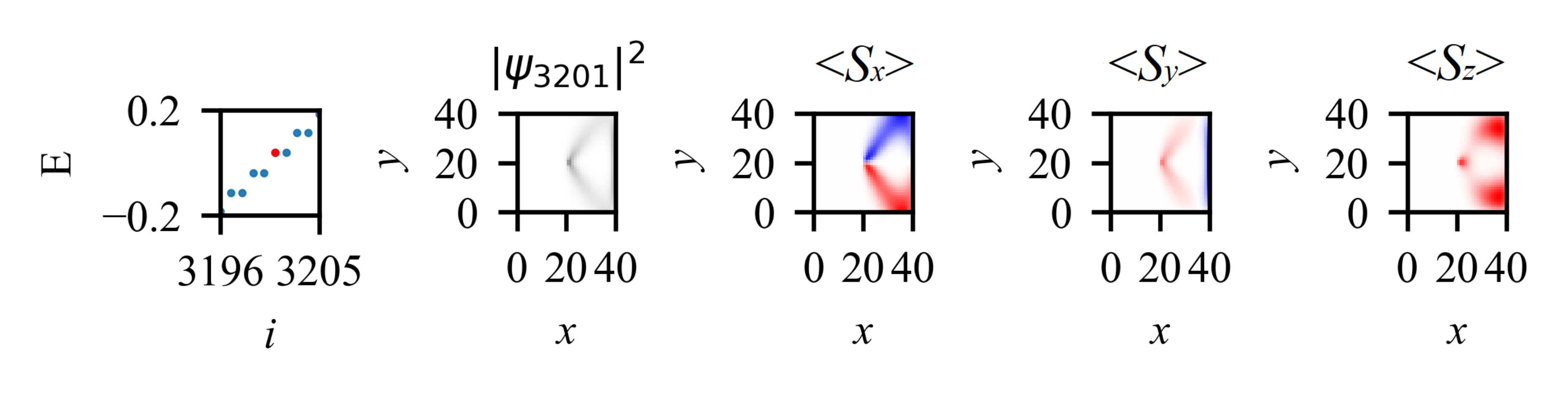}
    \includegraphics[width=1.0\textwidth]{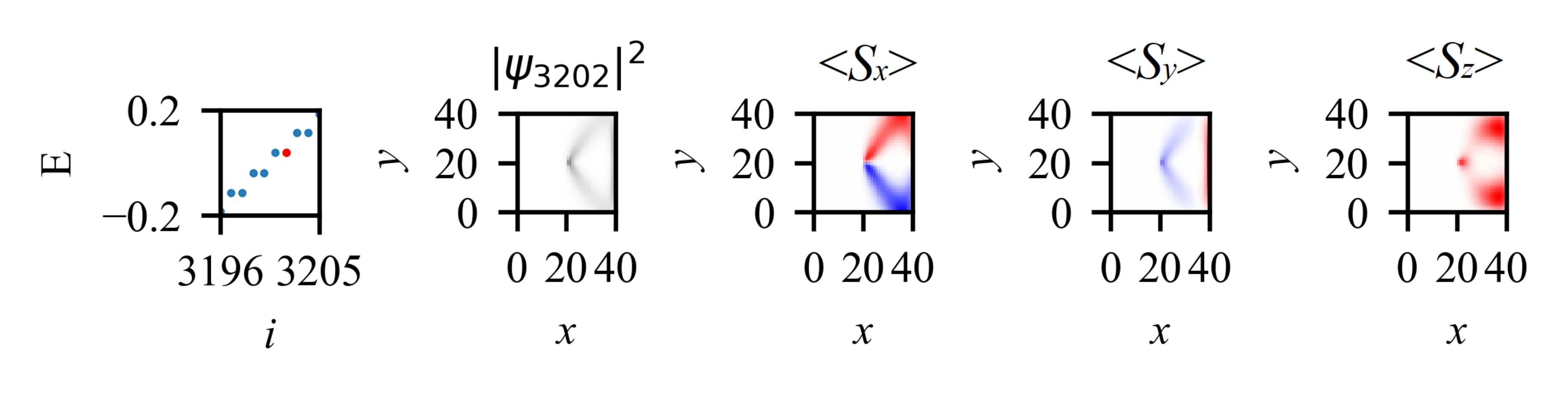}
    \caption{Spin texture of the edge states for the topological skyrmion semimetal defined over a square lattice geometry with point defect as depicted in Fig.~3 a); the first column consists of the energy spectrum where the plotted state is highlighted in red; the second column consists of the probability density distribution of the edge states, and the third, fourth and fifth columns are the different components of the spin expectation values $S_x$, $S_y$ and $S_x$. }
    \label{fig: spin texture}
\end{figure}

\newpage
The spin textures in the presence of an applied Zeeman field along the $+z$ axis of strength $B=0.1$ in units of energy, corresponding to $\mc{H}_z=B \tau_3\sigma_0$ added to the Hamiltonian Eq.~1, at the right vertical boundary of the system $x=L$:
\begin{figure}[h!]
    \centering
    \includegraphics[width=1.0\textwidth]{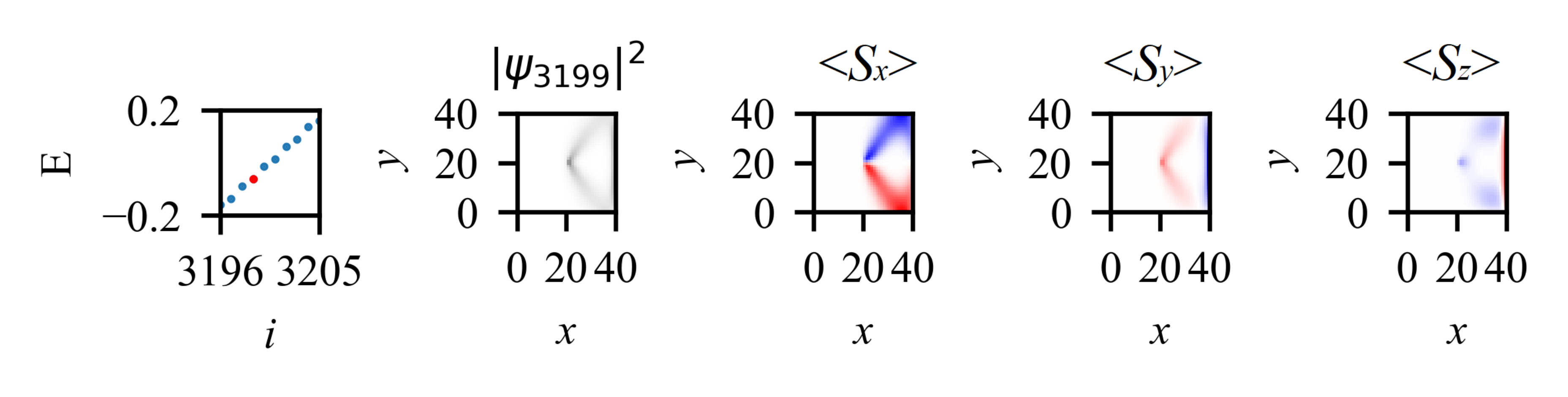}
    \includegraphics[width=1.0\textwidth]{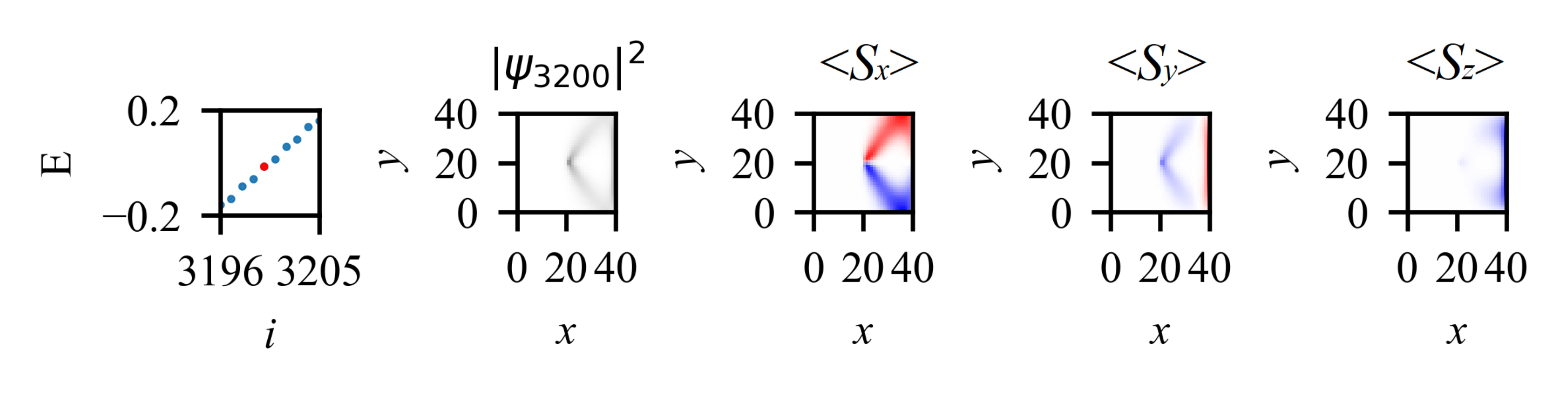}
    \includegraphics[width=1.0\textwidth]{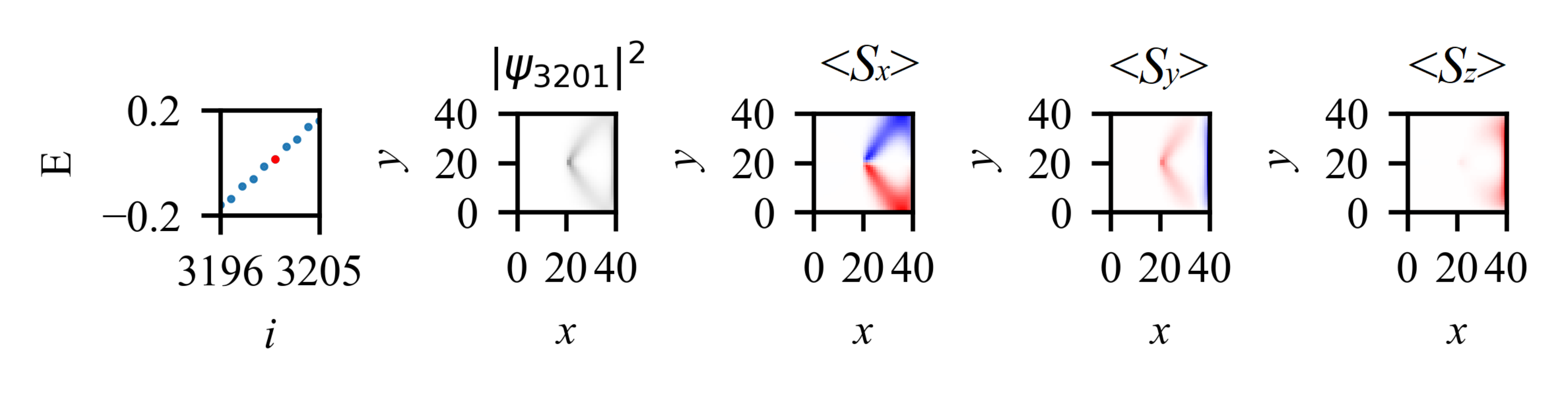}
    \includegraphics[width=1.0\textwidth]{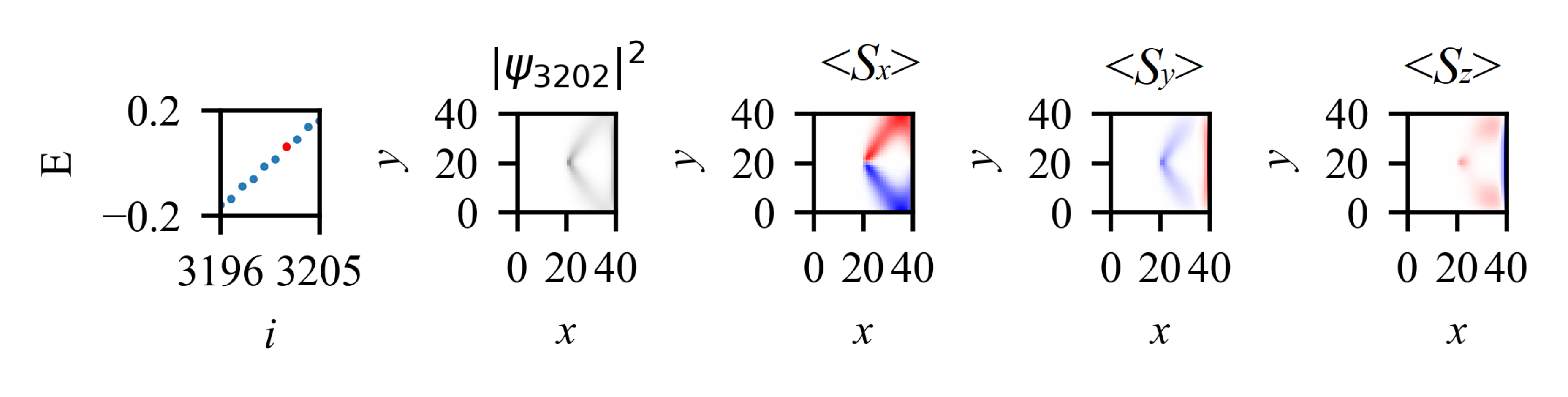}
    \caption{Spin texture of the edge states plotted in a similar manner to Fig.~\ref{fig: spin texture} with an external Zeeman magnetic field of strength $B=0.1$ in units of energy applied along the $+z$ axis corresponding to added term in the Hamiltonian $\mc{H}_z=B \tau_3\sigma_0$.}
    \label{fig: spin texture 2}
\end{figure}

\newpage
Here, we present the spin textures for the two-band Weyl semimetal given by $h(\boldsymbol{k})$ in the main text:
\begin{figure}[h!]
    \centering
    \includegraphics[width=1.0\textwidth]{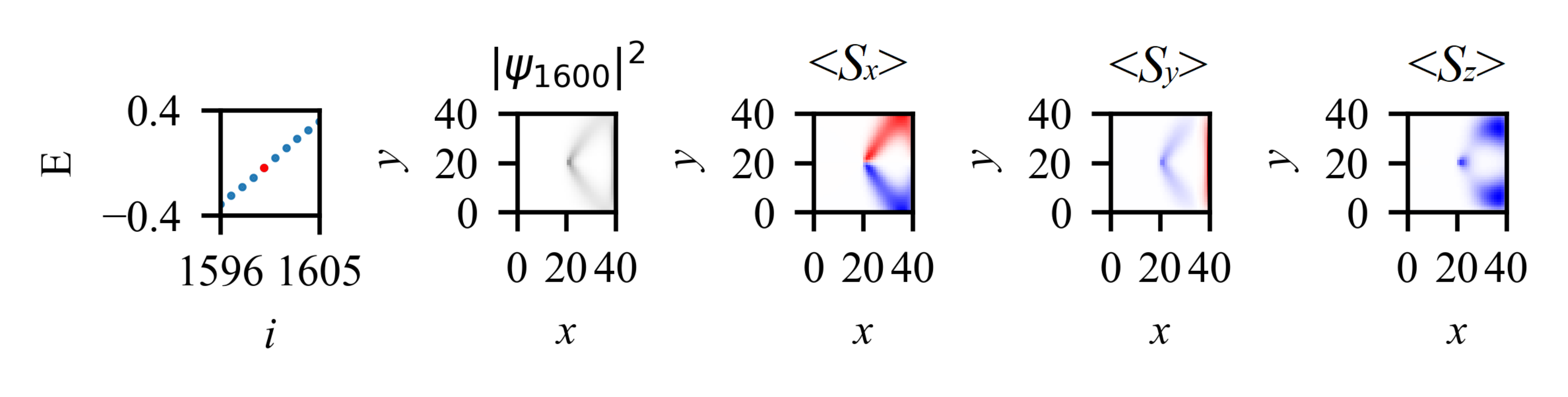}
    \includegraphics[width=1.0\textwidth]{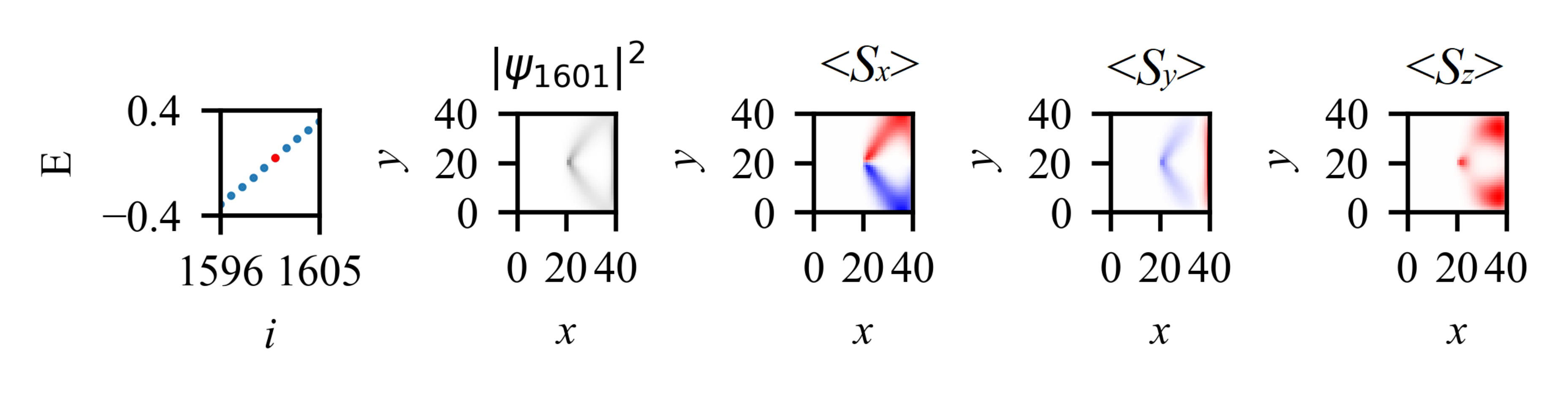}
    \caption{Spin texture of the edge states of the two-band Weyl semimetal, plotted in a similar manner to Fig.~\ref{fig: spin texture}.}
    \label{fig: spin texture 3}
\end{figure}

and also with applied Zeeman field term $h_z=B\sigma_0$:
\begin{figure}[h!]
    \centering
    \includegraphics[width=1.0\textwidth]{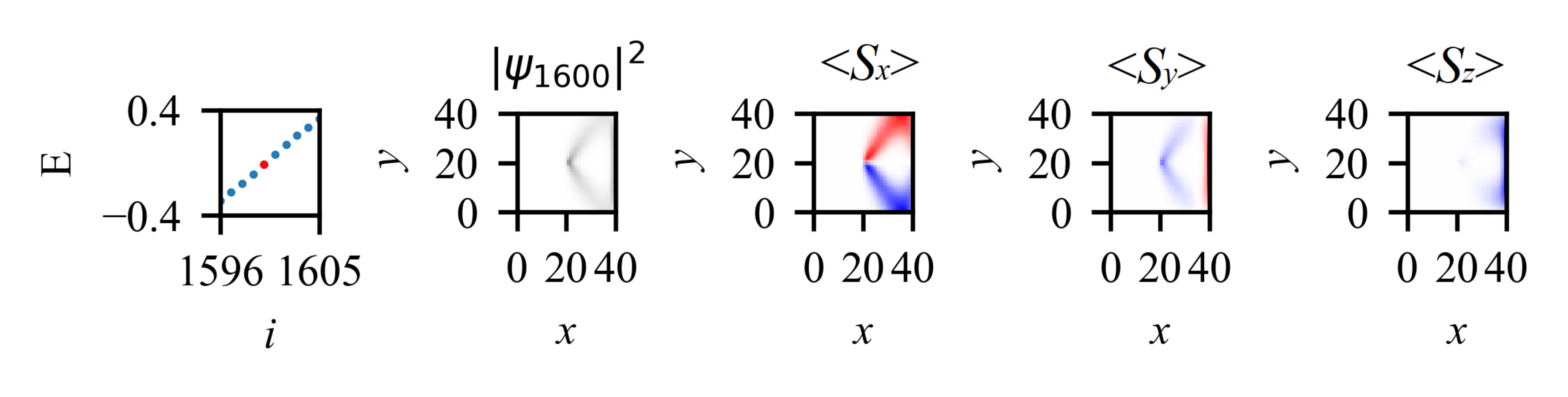}
    \includegraphics[width=1.0\textwidth]{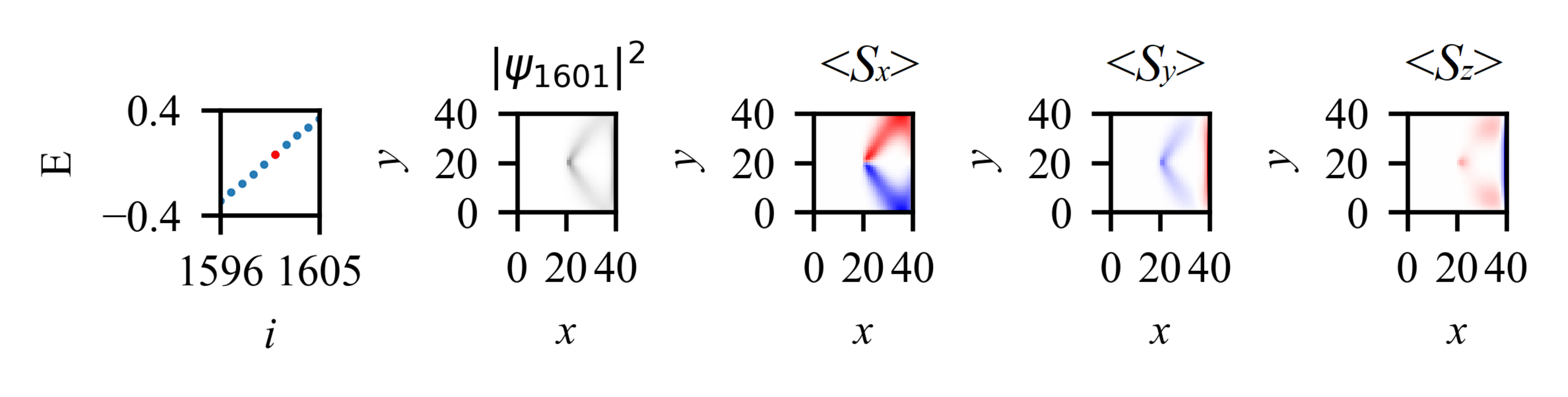}
    \caption{Spin texture of the edge states of a two-band Weyl semimetal given by $h(\boldsymbol{k})$ in the main text, with an additional applied magnetic field in the $+z$ axis corresponding to additional term in the Hamiltonian $h_z=B\sigma_0$, plotted in a similar manner to Fig.~\ref{fig: spin texture}.}
    \label{fig: spin texture 4}
\end{figure}

\newpage
\section{Details of three-band Bloch Hamiltonian for 2D chiral topological skyrmion phase}\label{Sup_sec_IV}
We first introduce three different embeddings of the Pauli matrices into $3 \times 3$ matrix representations $\boldsymbol{\sigma_\alpha}=(\sigma_{\alpha,x}, \sigma_{\alpha,y}, \sigma_{\alpha,z})$ where
\begin{align*}
\sigma_{\alpha,x} &= \begin{pmatrix}
0 & \delta_{\alpha1} & \delta_{\alpha2}\\
\delta_{\alpha1} & 0 & \delta_{\alpha3}\\
\delta_{\alpha2} & \delta_{\alpha3} & 0
\end{pmatrix},
\sigma_{\alpha,y} = i\begin{pmatrix}
0 & -\delta_{\alpha1} & -\delta_{\alpha2}\\
\delta_{\alpha1} & 0 & -\delta_{\alpha3}\\
\delta_{\alpha2} & \delta_{\alpha3} & 0
\end{pmatrix}\\
\sigma_{\alpha,z} &= \textnormal{diag}(\delta_{\alpha1}+\delta_{\alpha2},-\delta_{\alpha1},-\delta_{\alpha2}).
\label{3band spin matrices}
\end{align*}
Here, $\delta_{\alpha \beta} = 1$ for $\alpha = \beta$ and zero otherwise.

We may then write the Hamiltonian for the three-band topological skyrmion semimetal as $\mc{H} = \sum_{\boldsymbol{k}} \Psi^{\dagger}_{\boldsymbol{k}}\mc{H}\left(\boldsymbol{k}\right) \Psi^{}_{\boldsymbol{k}}$, where $\mc{H}\left(\boldsymbol{k}\right)$ takes the form of Eq.~7 in the main text with $\Psi_{\boldsymbol{k}} = \left( c_{\boldsymbol{k}, xy, \uparrow}, c_{\boldsymbol{k}, yz, \downarrow}, c_{\boldsymbol{k}, xz, \downarrow} \right)^{\top}$,
\begin{equation}
    \mathcal{H}(\bk) = \boldsymbol{d}_1(\bk) \cdot \boldsymbol{\sigma}_1 + \boldsymbol{d}_2(\bk) \cdot \boldsymbol{\sigma}_2 + \lambda \boldsymbol{\sigma}_{3,x}
    %\label{3band Ham},
\end{equation}
and the specific form of $\mc{H}\left(\boldsymbol{k}\right)$ relevant to Fig.~4 in the main text being
\begin{align*}
    d_{1,x}(\bk) &= 2\sin{(k_y)}\\
    d_{1,y}(\bk) &= 2\sin{(k_x)}\\
    d_{1,z}(\bk) &= m(k_z) - 2\cos{(k_y)} - 2\cos{(k_y)},
\end{align*}
with $m(k_z) = -1.5 - 0.4\cos{(k_z)}$, and
\begin{align*}
    d_{2,x}(\bk) &= 2\cos{(k_y)}\\
    d_{2,y}(\bk) &= 2\cos{(k_x)}\\
    d_{2,z}(\bk) &= m(k_z) - 2\sin{(k_y)} - 2\sin{(k_y)},
\end{align*}
and
\begin{equation*}
    d_{3,x}(\bk) = \lambda, \hspace{1pt}
    d_{3,y}(\bk) = 0, \hspace{12pt}
    d_{3,z}(\bk) = 0.
\end{equation*}
The spin expectation value is then computed using the following
spin operators introduced in past work\cite{cook2023, cook2023QSkHE}:
\begin{equation*}
    S_x = \frac{1}{2}\begin{pmatrix}
0 & 1 & 1\\
1 & 0 & 1\\
1 & 1 & 0
\end{pmatrix}, \hspace{12pt}
    S_y = \frac{1}{2}\begin{pmatrix}
0 & -i & -i\\
1 & 0 & -i\\
1 & 1 & 0
\end{pmatrix}, \hspace{12pt}
    S_z = \frac{1}{2}\begin{pmatrix}
2 & 0 & 0\\
0 & -1 & 0\\
0 & 0 & -1
\end{pmatrix}.
\end{equation*}

\end{document}